\documentclass[12pt,a4paper]{article}
\usepackage[a4paper, margin=1in]{geometry}
\usepackage[normalem]{ulem}
\usepackage{lmodern}
\usepackage[english]{babel}
\usepackage[utf8]{inputenc}
\usepackage[T1]{fontenc}
\usepackage{pdfpages}
\usepackage{microtype}
\usepackage{amsthm}
\usepackage{physics}
\usepackage{xcolor}
\usepackage{hyperref}
\usepackage{bm}
\usepackage{amsmath, amsfonts, amssymb}
\usepackage{authblk}
\usepackage{setspace}
\usepackage{framed}
\usepackage{cancel}
\usepackage{stackengine}
\stackMath
\hypersetup{colorlinks=true,linkcolor=blue,citecolor=blue,urlcolor=blue}

\begin{document}
	
	\begin{center}
		\textbf{{\LARGE Geometric structure of the  relativistic quantum phase space}}
	\end{center}
	
	\begin{center}
		\textbf{Philippe Manjakasoa Randriantsoa$ ^{1} $, Ravo Tokiniaina Ranaivoson$^{2}$,\\ Raoelina Andriambololona$^{3}$, Roland Raboanary$^{4}$, Wilfrid Chrysante Solofoarisina$^{5}$, Anjary Feno Hasina Rasamimanana$^{6}$.} \vspace{0.5cm}\\
	\end{center}
	
	\begin{center}
		\textit{ njakarandriantsoa@gmail.com}$ ^{1} $, 
		\textit{tokiniainaravor13@gmail.com}$ ^{2} $,
		\textit{raoelina.andriambololona@gmail.com}$ ^{3} $ ,
		\textit{r\_raboanary@yahoo.fr}$ ^{4} $, 
		\textit{wilfridc\_solofoarisina@yahoo.fr}$ ^{5} $, 
		\textit{anjaryhasinaetoile@gmail.com}$ ^{6} $ \vspace{0.5cm}\\
	\end{center}

	\begin{center}
		$ ^{1,2,3,5,6}$ Institut National des Sciences et Techniques Nucléaires (INSTN-Madagascar)\\
		BP 3907 Antananarivo 101, Madagascar,
		\textit{instn@moov.mg}\vspace{0.3cm}\\
		$ ^{1,2,3} ${\textit{TWAS Madagascar Chapter, Malagasy Academy,\\BP 4279 Antananarivo 101, Madagascar}}\vspace{0.3cm}\\
		$ ^{1,4, 6} ${\textit{ Faculty of Sciences, iHEPMAD-University of Antananarivo,\\ BP 566 Antananarivo 101, Madagascar}}\vspace{0.4cm}\\
		
		\textbf{Corresponding author: Philippe Manjakasoa Randriantsoa}	
		
	\end{center}
	
	\begin{abstract}
    The quest to reconcile quantum mechanics with gravitational theory motivates the exploration of frameworks that treat quantum uncertainty and spacetime geometry under a unified approach. A promising candidate that emerges from this pursuit is the relativistic quantum phase space (QPS) formalism, which extends classical phase space by incorporating both mean values and variance-covariance matrices of quantum states, providing a unified setting where the uncertainty principle and relativistic covariance coexist. For the signature $(1,4)$, we construct a scalar from the mean values and the inverse variance-covariance matrix and prove its invariance under linear canonical transformations (LCTs). Motivated by the form of the variance-covariance matrix in a particular reference frame, we identify this invariant as $\Gamma = L^2/\ell^2$ for states that saturate the uncertainty relations, where $L$ and $\ell$ are two fundamental length scales that can be identified with the de Sitter radius and the Planck length, respectively. From this invariant we obtain a geometric equation that unifies mean values and quantum fluctuations. In the limit $\ell \to 0$, the equation reduces to the de Sitter spacetime equation; in the limit $L \to \infty$, it yields a curved momentum space reminiscent of Born reciprocity. In the Minkowski limit (both $\ell \to 0$ and $L \to \infty$), the familiar relativistic relations for rest mass and proper time emerge. These limiting cases show how the Planck length and the cosmological constant can be unified within a single geometric constraint, establishing the QPS geometry as a promising framework for exploring the interplay between quantum mechanics and gravity.
	\end{abstract}

	
	    \newpage
	\section{Introduction}
	
	The quest for a consistent unification of quantum mechanics and general relativity remains one of the most profound challenges in theoretical physics. At the heart of this endeavor lies the tension between the uncertainty principle, which fundamentally limits the possibility to simultaneously determine particle momenta and coordinates, and the geometric structure of spacetime, which forms the backdrop of classical gravity. The concept of quantum phase space (QPS) offers a promising avenue to reconcile these seemingly incompatible notions by extending the classical phase space framework into the quantum domain while preserving both quantum and relativistic principles \cite{Ranaivoson2021, Ranaivoson2022, Ravelonjato2023}.
	
	In the QPS formalism, quantum states are described not only by their mean values $\langle p_{\mu} \rangle$ and $\langle x_{\mu} \rangle$ but also by their variance-covariance matrix, which encodes the intrinsic quantum fluctuations. The natural symmetry group of this construction is the group of linear canonical transformations (LCTs), which mix momenta and coordinates operators while preserving the canonical commutation relations \cite{Ranaivoson2021, Ranaivoson2022, Andriambololona2017, Randriantsoa2025, Ranaivoson2025}. For a spacetime with signature $(N_{+}, N_{-})$, the LCT group is isomorphic to the symplectic group $\mathrm{Sp}(2N_{+}, 2N_{-})$.
	
	A particularly interesting case arises for the signature $(1,4)$, which corresponds to the de Sitter group $\mathrm{SO}(1,4)$. This choice is physically significant for two reasons. First, de Sitter spacetime describes a universe with a positive cosmological constant, consistent with current cosmological observations \cite{Schneider2023, Heavens2008, LopezCorredoira2017, GenovaSantos2020}. Second, Refs.~\cite{Andriambololona2021, Ranaivoson2025, Randriantsoa2025} have shown that the same symplectic group $\mathrm{Sp}(2,8)$ leads to a classification of quarks and leptons that includes sterile neutrinos. In the present work, we focus on the geometric structure of the quantum phase space itself, deriving a scalar invariant and a unified geometric equation that incorporates both the Planck length and the de Sitter radius. Our results are compatible with those particle-physics predictions but are independent of them.
	
	In this work, we examine the basic geometric properties of the quantum phase space and their physical consequences. We construct a scalar invariant $\Gamma$ from the mean values and the inverse variance-covariance matrix, demonstrating its invariance under LCTs. By considering quantum states that saturate the uncertainty relations---those that are as close as possible to classical phase space points and which correspond to the definition of the QPS itself \cite{Ranaivoson2022, Ravelonjato2023, Ranaivoson2025, Randriantsoa2025}---we show that $\Gamma$ encodes two fundamental length scales: a scale $L$ that can be identified with the de Sitter radius (maximal coordinate uncertainty) and a scale $\ell$ that can be identified with the Planck length (minimal coordinate uncertainty). From the invariance of $\Gamma$ we derive a geometric equation that unifies the mean values and the quantum fluctuations. Analyzing two asymptotic regimes of this equation leads, respectively, to the de Sitter spacetime equation and a de Sitter-like structure in momentum space. Moreover, we show how the Poincaré limit of the geometric equation naturally gives rise to the concepts of rest mass and proper time, thereby providing a geometric origin for these fundamental quantities.
	
	The paper is organized as follows. Section 2 analyzes the basic geometry of the quantum phase space and derives the scalar invariant $\Gamma$ associated with LCTs. Section 3 establishes the geometric equation of the quantum phase space and examines the two limits that lead to de Sitter spacetime and curved de Sitter-like momentum space. Section 4 demonstrates how rest mass and proper time emerge from the geometry in the classical limit. Section 5 presents a discussion of the results, highlighting connections to Born reciprocity, particle physics and cosmology, and outlines directions for future research. The notations used are inspired by the indications in the reference \cite{Andriambololona1985}, and bold letters are generally used to denote operators.
	
	\section{Scalar invariant associated with LCTs}
	
	The relativistic quantum phase space formalism is built upon the canonical commutation relations (CCRs) generalized to a spacetime with signature $(N_{+}, N_{-})$ and dimension $N = N_{+} + N_{-}$ \cite{Ranaivoson2021, Ranaivoson2022}. The CCRs are given by:
	\begin{equation}
		\begin{cases}
			[\mathbf{p}_{\mu}, \mathbf{x}_{\nu}] = i\hbar \eta_{\mu\nu},\\
			[\mathbf{p}_{\mu}, \mathbf{p}_{\nu}] = 0,\\
			[\mathbf{x}_{\mu}, \mathbf{x}_{\nu}] = 0,
		\end{cases}
	\end{equation}
	where $\eta_{\mu\nu} = \mathrm{diag}(1, \ldots, 1, -1, \ldots, -1)$ with $N_{+}$ positive entries and $N_{-}$ negative entries and $\hbar$ is the Planck reduced constant.
	
	A linear canonical transformation (LCT) is a linear transformation mixing the momenta and coordinates operators $(\mathbf{p}_{\mu}$ and $\mathbf{x}_{\mu})$ that preserves the CCRs \cite{Ranaivoson2021, Ranaivoson2022, Andriambololona2017}:
	\begin{equation}
		\begin{cases}
			\mathbf{p}'_{\mu} = A^{\nu}_{\mu}\mathbf{p}_{\nu} + \dfrac{\hbar}{L^2} B^{\nu}_{\mu}\mathbf{x}_{\nu},\\[8pt]
			\mathbf{x}'_{\mu} = \dfrac{\ell^2}{\hbar} C^{\nu}_{\mu}\mathbf{p}_{\nu} + D^{\nu}_{\mu}\mathbf{x}_{\nu},\\[8pt]
			[\mathbf{p}'_{\mu}, \mathbf{x}'_{\nu}] = [\mathbf{p}_{\mu}, \mathbf{x}_{\nu}] = i\hbar \eta_{\mu\nu},\\
			[\mathbf{p}'_{\mu}, \mathbf{p}'_{\nu}] = [\mathbf{p}_{\mu}, \mathbf{p}_{\nu}] = 0,\\
			[\mathbf{x}'_{\mu}, \mathbf{x}'_{\nu}] = [\mathbf{x}_{\mu}, \mathbf{x}_{\nu}] = 0.
		\end{cases}
	\end{equation}
	
	The factors involving the parameters $\ell$ and $L$ ensure that the coefficients $A^{\nu}_{\mu}$, $B^{\nu}_{\mu}$, $C^{\nu}_{\mu}$, $D^{\nu}_{\mu}$ appearing in the linear canonical transformations are dimensionless. They can be interpreted as the minimal and maximal possible values of coordinates uncertainties. In this context, they may be identified respectively with the Planck length $\ell_P = \sqrt{\hbar G/c^3}$ and the de Sitter radius $R_{dS} = \sqrt{3/\Lambda}$, where $G$ is the gravitational constant, $c$ is the speed of light in vacuum, and $\Lambda$ is the cosmological constant \cite{Ranaivoson2021, Schneider2023, Meschini2007}.
	
	If we denote respectively $A$, $B$, $C$, $D$ the $N \times N$ matrices associated to the coefficients $A^{\nu}_{\mu}$, $B^{\nu}_{\mu}$, $C^{\nu}_{\mu}$, $D^{\nu}_{\mu}$, it can be deduced from the definition of LCT that the matrix
	\[
	\begin{pmatrix}
		A & \dfrac{\hbar}{L^2}B \\[6pt]
		\dfrac{\ell^2}{\hbar}C & D
	\end{pmatrix}
	\]
	belongs to the symplectic group $\mathrm{Sp}(2N_{+}, 2N_{-})$; i.e., it satisfies the following relation:
	\begin{equation}
		\begin{pmatrix}
			A & \dfrac{\ell^2}{\hbar}C \\[6pt]
			\dfrac{\hbar}{L^2}B & D
		\end{pmatrix}^{T}
		\begin{pmatrix}
			0 & \eta \\
			-\eta & 0
		\end{pmatrix}
		\begin{pmatrix}
			A & \dfrac{\ell^2}{\hbar}C \\[6pt]
			\dfrac{\hbar}{L^2}B & D
		\end{pmatrix}
		=
		\begin{pmatrix}
			0 & \eta \\
			-\eta & 0
		\end{pmatrix},
	\end{equation}
	with $\eta = \mathrm{diag}(\underbrace{1, \ldots, 1}_{N_{+}}, \underbrace{-1, \ldots, -1}_{N_{-}})$.
	
	For the specific case of signature $(1,4)$ considered in this work, the LCT group is isomorphic to $\mathrm{Sp}(2,8)$ \cite{Ranaivoson2021, Randriantsoa2025}.
	
	Let $|\psi\rangle$ be a quantum state. We define the mean values and variance-covariance matrices \cite{Ranaivoson2022, Ranaivoson2025}:
	\begin{equation}
		\begin{cases}
			\langle \mathbf{p}_{\mu} \rangle = \langle \psi | \mathbf{p}_{\mu} | \psi \rangle,\\
			\langle \mathbf{x}_{\mu} \rangle = \langle \psi | \mathbf{x}_{\mu} | \psi \rangle,\\
			\mathcal{P}_{\mu\nu} = \langle \psi | (\mathbf{p}_{\mu} - \langle \mathbf{p}_{\mu} \rangle)(\mathbf{p}_{\nu} - \langle \mathbf{p}_{\nu} \rangle) | \psi \rangle,\\
			\mathcal{X}_{\mu\nu} = \langle \psi | (\mathbf{x}_{\mu} - \langle \mathbf{x}_{\mu} \rangle)(\mathbf{x}_{\nu} - \langle \mathbf{x}_{\nu} \rangle) | \psi \rangle,\\
			\mathcal{Q}_{\mu\nu} = \frac{1}{2} \langle \psi | (\mathbf{p}_{\mu} - \langle \mathbf{p}_{\mu} \rangle)(\mathbf{x}_{\nu} - \langle \mathbf{x}_{\nu} \rangle) + (\mathbf{x}_{\nu} - \langle \mathbf{x}_{\nu} \rangle)(\mathbf{p}_{\mu} - \langle \mathbf{p}_{\mu} \rangle) | \psi \rangle.
		\end{cases}
	\end{equation}
	
	Under an LCT, the mean values transform as \cite{Ranaivoson2022}:
	\begin{equation}
		\begin{pmatrix}
			\langle \mathbf{p}' \rangle \\
			\langle \mathbf{x}' \rangle
		\end{pmatrix}
		=
		\begin{pmatrix}
			A & \dfrac{\ell^2}{\hbar}C \\[6pt]
			\dfrac{\hbar}{L^2}B & D
		\end{pmatrix}
		\begin{pmatrix}
			\langle \mathbf{p} \rangle \\
			\langle \mathbf{x} \rangle
		\end{pmatrix},
	\end{equation}
	where we have introduced the fundamental scales $L$ and $\ell$ associated with the quantum phase space \cite{Ranaivoson2021}. The variance-covariance matrix transforms as:
	\begin{equation}
		\begin{pmatrix}
			\mathcal{P}' & \mathcal{Q}' \\
			\mathcal{Q}'^T & \mathcal{X}'
		\end{pmatrix}
		=
		\begin{pmatrix}
			A & \dfrac{\ell^2}{\hbar}C \\[6pt]
			\dfrac{\hbar}{L^2}B & D
		\end{pmatrix}^{T}
		\begin{pmatrix}
			\mathcal{P} & \mathcal{Q} \\
			\mathcal{Q}^T & \mathcal{X}
		\end{pmatrix}
		\begin{pmatrix}
			A & \dfrac{\ell^2}{\hbar}C \\[6pt]
			\dfrac{\hbar}{L^2}B & D
		\end{pmatrix}.
	\end{equation}
	
	From these transformation laws, we can construct a scalar invariant. Consider the quantity:
	\begin{equation}
		\Gamma =
		\begin{pmatrix}
			\langle \mathbf{p} \rangle & \langle \mathbf{x} \rangle
		\end{pmatrix}
		\begin{pmatrix}
			\mathcal{P} & \mathcal{Q} \\
			\mathcal{Q}^T & \mathcal{X}
		\end{pmatrix}^{-1}
		\begin{pmatrix}
			\langle \mathbf{p} \rangle \\
			\langle \mathbf{x} \rangle
		\end{pmatrix}.
	\end{equation}
	
	Using the transformation properties, one can verify that $\Gamma$ is invariant under LCTs. Indeed:
	\begin{equation}
		\Gamma' =
		\begin{pmatrix}
			\langle \mathbf{p}' \rangle & \langle \mathbf{x}' \rangle
		\end{pmatrix}
		\begin{pmatrix}
			\mathcal{P}' & \mathcal{Q}' \\
			\mathcal{Q}'^T & \mathcal{X}'
		\end{pmatrix}^{-1}
		\begin{pmatrix}
			\langle \mathbf{p}' \rangle \\
			\langle \mathbf{x}' \rangle
		\end{pmatrix}
	\end{equation}
	\begin{equation}
		=
		\begin{pmatrix}
			\langle \mathbf{p} \rangle & \langle \mathbf{x} \rangle
		\end{pmatrix}
		M
		\left[
		M^{-1}
		\begin{pmatrix}
			\mathcal{P} & \mathcal{Q} \\
			\mathcal{Q}^T & \mathcal{X}
		\end{pmatrix}^{-1}
		(M^T)^{-1}
		\right]
		M^T
		\begin{pmatrix}
			\langle \mathbf{p} \rangle \\
			\langle \mathbf{x} \rangle
		\end{pmatrix}
		= \Gamma,
	\end{equation}
	where $M =
	\begin{pmatrix}
		A & \dfrac{\ell^2}{\hbar}C \\[6pt]
		\dfrac{\hbar}{L^2}B & D
	\end{pmatrix}$.
	
	Thus, $\Gamma$ is a scalar invariant associated to linear canonical transformations.
	
	\section{Geometric equation of the quantum phase space}
	
	In this section, our goal is to derive the geometric equation of the quantum phase space, from which a quantum de Sitter spacetime equation can be obtained in the limit $\ell \to 0$.
	
	The quantum phase space is defined as the set $\{(\langle \mathbf{p}_{\mu} \rangle, \langle \mathbf{x}_{\mu} \rangle)\}$ of the mean values for a given values of the variance-covariance matrix $\begin{pmatrix} \mathcal{P} & \mathcal{Q} \\ \mathcal{Q}^{T} & \mathcal{X} \end{pmatrix}$ which saturate the uncertainty relation \cite{Ranaivoson2022, Ravelonjato2023, Ranaivoson2025, Randriantsoa2025}, i.e., we have the following relation:
	\begin{equation}
		\mathcal{P}_{\mu\mu}\mathcal{X}_{\mu\mu} - (\mathcal{Q}_{\mu\mu})^2 = \frac{\hbar^2}{4}.
	\end{equation}
	
	The saturation equation (11) corresponds to the quantum state $|\psi\rangle = |\langle z\rangle \rangle$ that are the most similar to classical phase space point state. The wave-function $\langle \mathbf{x}_{\mu}|\langle z\rangle \rangle$ associated to these states $|z\rangle$ are Gaussian-like function \cite{Ranaivoson2021, Ranaivoson2022, Ranaivoson2025}:
	\begin{equation}
		\langle \mathbf{x}^{\mu}|\{\langle z_{\mu}\rangle \} \rangle = \langle \mathbf{x}|\langle z\rangle \rangle = 
		\frac{e^{-\frac{B_{\mu\nu}}{\hbar^2}(x^{\mu} - \langle x^{\mu}\rangle)(x^{\nu} - \langle x^{\nu}\rangle) - \frac{i}{\hbar}\langle p_{\mu}\rangle x^{\mu} + iK}}{|\langle 2\pi\rangle^N|\det[\mathcal{X}_{\mu}^{\mu}]|^{1/4}},
	\end{equation}
	
	We define the geometric equation of the quantum phase space as a relation connecting the mean values $\langle \mathbf{p}_{\mu}\rangle$ and $\langle \mathbf{x}_{\mu}\rangle$ associated with the quantum states $|z\rangle$ for $\mu = 0,1,2,3,4$. This equation is assumed to admit the following limits:
	
	The quantum spacetime de Sitter limit $(\ell \to 0)$:
	\begin{equation}
		(\langle x_0\rangle)^2 - (\langle x_1\rangle)^2 - (\langle x_2\rangle)^2 - (\langle x_3\rangle)^2 - (\langle x_4\rangle)^2 = -L^2.
	\end{equation}
	
	The quantum momentum space limit $(L \to \infty)$:
	\begin{equation}
		(\langle p_0\rangle)^2 - (\langle p_1\rangle)^2 - (\langle p_2\rangle)^2 - (\langle p_3\rangle)^2 - (\langle p_4\rangle)^2 = -\left(\frac{\hbar}{2\ell}\right)^2.
	\end{equation}
	
	Equation (13) corresponds to the identification of $L$ with the de Sitter radius. Similarly, $\ell$ may be identified with the Planck length, and $\frac{\hbar}{2\ell}$ may be then interpreted as the maximum value of the momentum uncertainties.
	
	Two particular elements belonging to the spaces defined by the relations (13) and (14) satisfy the following relation:
	\begin{equation}
		\begin{cases}
			\langle x_0\rangle = \langle x_1\rangle = \langle x_2\rangle = \langle x_3\rangle = 0, & \langle x_4\rangle = L,\\
			\langle p_0\rangle = \langle p_1\rangle = \langle p_2\rangle = \langle p_3\rangle = 0, & \langle p_4\rangle = \dfrac{\hbar}{2\ell}.
		\end{cases}
	\end{equation}
	
	Any elements of the respective space can be deduced from these two elements by using the action of the de Sitter group which can be viewed as particular LCTs corresponding to the following relation:
	\begin{equation}
		\begin{pmatrix}
			\langle \mathbf{p}' \rangle \\
			\langle \mathbf{x}' \rangle
		\end{pmatrix}
		=
		\begin{pmatrix}
			\mathbb{A} & 0 \\
			0 & \mathbb{A}
		\end{pmatrix}
		\begin{pmatrix}
			\langle \mathbf{p} \rangle \\
			\langle \mathbf{x} \rangle
		\end{pmatrix}.
	\end{equation}
	
	Taking into account the relation (3), it is easy to verify that $\mathbb{A}$ satisfies the relation $\mathbb{A}^{T}\eta \mathbb{A} = \eta$ and $\det(\mathbb{A}) = 1$, i.e., belongs to the de Sitter group $\mathrm{SO}(1,4)$. As discussed in Refs.~\cite{Ranaivoson2021, Ranaivoson2022}, an LCT can be interpreted as a change of reference frame that transforms the momenta and coordinates operators while leaving the quantum state unchanged. Under such a transformation, the mean values and variance-covariance matrices of the momenta and coordinates operators are also modified. Within this framework, the values considered in Eqs.~(15) can be interpreted as the mean values corresponding to a state $|\langle z \rangle \rangle$ in a particular reference frame.
	
	For the general case $(\ell \neq 0$ and $L$ finite), instead of the relation (15) we can just choose a particular reference $\mathcal{F}_0$ in which we have the following relation:
	\begin{equation}
		\begin{cases}
			\langle x_0\rangle_0 = \langle x_1\rangle_0 = \langle x_2\rangle_0 = \langle x_3\rangle_0 = 0, & \langle x_4\rangle_0 = \lambda,\\
			\langle p_0\rangle_0 = \langle p_1\rangle_0 = \langle p_2\rangle_0 = \langle p_3\rangle_0 = 0, & \langle p_4\rangle_0 = \kappa.
		\end{cases}
	\end{equation}
	
	We may also suppose that the variance-covariance values associated with this state in this particular frame and which satisfy the uncertainties saturation condition in Eqs.~(11) are:
	\begin{equation}
		\begin{cases}
			\mathcal{P}_{\mu\nu} = \delta_{\mu\nu}\dfrac{\hbar^2}{4\ell^2},\\
			\mathcal{X}_{\mu\nu} = \delta_{\mu\nu} L^2,\\
			\mathcal{Q}_{\mu\nu} = \delta_{\mu\nu}\dfrac{\hbar}{2\ell}\sqrt{L^2 - \ell^2}.
		\end{cases}
	\end{equation}
	
	In this frame, the variance-covariance matrix of the considered quantum state takes then the following block-diagonal form:
	\begin{equation}
		\begin{pmatrix}
			\mathcal{P} & \mathcal{Q}\\
			\mathcal{Q}^T & \mathcal{X}
		\end{pmatrix}
		=
		\begin{pmatrix}
			\dfrac{\hbar^2}{4\ell^2} I_5 & \dfrac{\hbar}{2\ell}\sqrt{L^2 - \ell^2} I_5\\[8pt]
			\dfrac{\hbar}{2\ell}\sqrt{L^2 - \ell^2} I_5 & L^2 I_5
		\end{pmatrix},
	\end{equation}
	where $I_5$ is the $5 \times 5$ identity matrix. Its inverse is:
	\begin{equation}
		\begin{pmatrix}
			\mathcal{P} & \mathcal{Q}\\
			\mathcal{Q}^T & \mathcal{X}
		\end{pmatrix}^{-1}
		=
		\begin{pmatrix}
			\dfrac{4L^2}{\hbar^2} I_5 & -\dfrac{2}{\ell\hbar}\sqrt{L^2 - \ell^2} I_5\\[8pt]
			-\dfrac{2}{\ell\hbar}\sqrt{L^2 - \ell^2} I_5 & \dfrac{1}{\ell^2} I_5
		\end{pmatrix}.
	\end{equation}
	
	Substituting the mean values into the expression for $\Gamma$, we obtain:
	\begin{equation}
		\frac{4L^2}{\hbar^2}\kappa^2 - \frac{4}{\ell\hbar}\sqrt{L^2 - \ell^2}\,\kappa\lambda + \frac{1}{\ell^2}\lambda^2 = \Gamma.
	\end{equation}
	
	We may consider another reference of frame $\mathcal{F}$ related to $\mathcal{F}_0$ by a de Sitter transformation defined by a matrix $\mathbb{A}$ (and its inverse $\mathbb{M} = \mathbb{A}^{-1}$):
	\begin{equation}
		\begin{cases}
			\langle p_{\mu}\rangle = \mathbb{A}_{\mu}^{\nu}\langle p_{\mu}\rangle_0 = \mathbb{A}_{\mu}^{4}\kappa,\\
			\langle x_{\mu}\rangle = \mathbb{A}_{\mu}^{\nu}\langle x_{\mu}\rangle_0 = \mathbb{A}_{\mu}^{4}\lambda,
		\end{cases}
		\iff
		\begin{cases}
			\kappa = \langle p_4\rangle_0 = \mathbb{M}_{4}^{\mu}\langle p_{\mu}\rangle,\\
			\lambda = \langle x_4\rangle_0 = \mathbb{M}_{4}^{\nu}\langle x_{\nu}\rangle.
		\end{cases}
	\end{equation}
	
	In the frame $\mathcal{F}$, Eq.~(21) becomes ($\Gamma$ is invariant):
	\begin{equation}
		-\frac{4L^2}{\hbar^2}\eta_{\mu\nu}\langle p_{\mu}\rangle \langle p_{\nu}\rangle
		-\frac{4}{\ell\hbar}\sqrt{L^2 - \ell^2}\,\mathbb{M}_{4}^{\mu}\mathbb{M}_{4}^{\nu}\langle p_{\mu}\rangle \langle x_{\nu}\rangle
		-\frac{1}{\ell^2}\eta_{\mu\nu}\langle x_{\mu}\rangle \langle x_{\nu}\rangle = \Gamma.
	\end{equation}
	
	To obtain Eq.~(23), we have used the fact that for the de Sitter subgroup, $\mathbb{M}_{4}^{\mu}\mathbb{M}_{4}^{\nu} = \eta^{\mu\nu}$ (up to a sign) because the fourth row of $\mathbb{M}$ forms a time-like vector. This follows directly from the de Sitter condition $\mathbb{M}^{T}\eta\mathbb{M} = \eta$, which implies $\mathbb{M}_{4}^{\mu}\mathbb{M}_{4}^{\nu}\eta_{\mu\nu} = \eta_{44} = -1$.
	
	Equation (23) can be put in the following form:
	\begin{equation}
		\frac{1}{\Gamma}
		\left[
		-\frac{4L^2}{\hbar^2}\eta_{\mu\nu}\langle p_{\mu}\rangle \langle p_{\nu}\rangle
		-\frac{4}{\ell\hbar}\sqrt{L^2 - \ell^2}\,\mathbb{M}_{4}^{\mu}\mathbb{M}_{4}^{\nu}\langle p_{\mu}\rangle \langle x_{\nu}\rangle
		-\frac{1}{\ell^2}\eta_{\mu\nu}\langle x_{\mu}\rangle \langle x_{\nu}\rangle
		\right] = 1.
	\end{equation}
	
	The Eqs.~(13) and (14) can be deduced respectively from Eqs.~(24) if we choose:
	\begin{equation}
		\Gamma = \frac{L^2}{\ell^2}.
	\end{equation}
	
	This choice is justified as follows: For the class of Gaussian states that saturate the uncertainty relations and have the variance-covariance matrix (19)-(20) in the special frame $\mathcal{F}_0$, a direct substitution into the definition of $\Gamma$ yields $\Gamma = L^2/\ell^2$. Indeed, using the inverse variance-covariance matrix (20) and the mean values (17), we obtain:
	\[
	\Gamma = \frac{4L^2}{\hbar^2}\kappa^2 - \frac{4}{\ell\hbar}\sqrt{L^2-\ell^2}\,\kappa\lambda + \frac{1}{\ell^2}\lambda^2.
	\]
	For the states saturating the uncertainty relations in the frame $\mathcal{F}_0$, we have $\kappa = \hbar/(2\ell)$ and $\lambda = L$, which gives:
	\[
	\Gamma = \frac{4L^2}{\hbar^2}\frac{\hbar^2}{4\ell^2} - \frac{4}{\ell\hbar}\sqrt{L^2-\ell^2}\frac{\hbar}{2\ell}L + \frac{1}{\ell^2}L^2
	= \frac{L^2}{\ell^2} - \frac{2L}{\ell^2}\sqrt{L^2-\ell^2} + \frac{L^2}{\ell^2}
	= \frac{L^2}{\ell^2}.
	\]
	Thus the value $\Gamma = L^2/\ell^2$ is not arbitrary but follows from the saturation of the uncertainty relations and the specific form of the variance-covariance matrix in the special frame.
	
	With this choice, Eq.~(24) becomes:
	\begin{equation}
		-\frac{4\ell^2}{\hbar^2}\eta_{\mu\nu}\langle p_{\mu}\rangle \langle p_{\nu}\rangle
		-\frac{4\ell}{\hbar L^2}\sqrt{L^2 - \ell^2}\,\mathbb{M}_{4}^{\mu}\mathbb{M}_{4}^{\nu}\langle p_{\mu}\rangle \langle x_{\nu}\rangle
		-\frac{1}{L^2}\eta_{\mu\nu}\langle x_{\mu}\rangle \langle x_{\nu}\rangle = 1.
	\end{equation}
	
	And we have explicitly the following limits:
	
	For $\ell \to 0$, Eq.~(26) becomes:
	\begin{equation}
		-\frac{1}{L^2}\eta_{\mu\nu}\langle x_{\mu}\rangle \langle x_{\nu}\rangle = 1
		\Longrightarrow
		(\langle x_0\rangle)^2 - (\langle x_1\rangle)^2 - (\langle x_2\rangle)^2 - (\langle x_3\rangle)^2 - (\langle x_4\rangle)^2 = -L^2.
	\end{equation}
	
	Eq.~(27) corresponds exactly to the quantum de Sitter spacetime equation (13).
	
	For $L \to \infty$, Eq.~(26) becomes:
	\begin{equation}
		-\frac{4\ell^2}{\hbar^2}\eta_{\mu\nu}\langle p_{\mu}\rangle \langle p_{\nu}\rangle = 1
		\Longrightarrow
		(\langle p_0\rangle)^2 - (\langle p_1\rangle)^2 - (\langle p_2\rangle)^2 - (\langle p_3\rangle)^2 - (\langle p_4\rangle)^2 = -\left(\frac{\hbar}{2\ell}\right)^2.
	\end{equation}
	
	Equation (28) is precisely the quantum momenta space equation (14).
	
	The most general form of the geometric equation of the quantum phase space can be deduced from the Eqs.~(7) and (25):
	\begin{equation}
		\begin{pmatrix}
			\langle \mathbf{p} \rangle & \langle \mathbf{x} \rangle
		\end{pmatrix}
		\begin{pmatrix}
			\mathcal{P} & \mathcal{Q} \\
			\mathcal{Q}^T & \mathcal{X}
		\end{pmatrix}^{-1}
		\begin{pmatrix}
			\langle \mathbf{p} \rangle \\
			\langle \mathbf{x} \rangle
		\end{pmatrix}
		= \frac{L^2}{\ell^2}.
	\end{equation}
	
	\section{Emergence of mass and spacetime from quantum phase space geometry}
	
	The geometric equation derived in the previous section,
	\begin{equation}
		\begin{pmatrix}
			\langle \mathbf{p} \rangle & \langle \mathbf{x} \rangle
		\end{pmatrix}
		\begin{pmatrix}
			\mathcal{P} & \mathcal{Q} \\
			\mathcal{Q}^T & \mathcal{X}
		\end{pmatrix}^{-1}
		\begin{pmatrix}
			\langle \mathbf{p} \rangle \\
			\langle \mathbf{x} \rangle
		\end{pmatrix}
		= \frac{L^2}{\ell^2},
	\end{equation}
	provides a unified constraint linking the mean values and quantum fluctuations. While the limits $\ell \to 0$ and $L \to \infty$ recover de Sitter spacetime and a curved momentum space respectively, a more fundamental physical interpretation emerges when we consider specific reference frames and identify the physical meaning of the fifth coordinate $x_4$ and momentum $p_4$.
	
	\subsection{The fifth dimension and the special case of vanishing coordinates-momenta correlations}
	
	In the particular frame $\mathcal{F}_0$ introduced in Sec.~3, the mean values are aligned along the fourth direction:
	\begin{equation}
		\begin{cases}
			\langle x_0\rangle = \langle x_1\rangle = \langle x_2\rangle = \langle x_3\rangle = 0, & \langle x_4\rangle = L,\\
			\langle p_0\rangle = \langle p_1\rangle = \langle p_2\rangle = \langle p_3\rangle = 0, & \langle p_4\rangle = \dfrac{\hbar}{2\ell}.
		\end{cases}
	\end{equation}
	
	In this frame, the fifth components $\langle x_4\rangle$ and $\langle p_4\rangle$ are not merely auxiliary; they encode the fundamental curvature scales of the quantum phase space.
	
	\subsection{Simplified geometric relation for $\mathcal{Q} = 0$}
	
	For a different family of quantum states, we may consider the special case where the coordinates-momenta correlations vanish, i.e., $\mathcal{Q}_{\mu\nu} = 0$. This corresponds to a simpler class of states with no position-momentum correlations. In this case, the variance-covariance matrices are maximal and isotropic. In this frame:
	\begin{equation}
		\begin{pmatrix}
			\mathcal{P} & \mathcal{Q} \\
			\mathcal{Q}^T & \mathcal{X}
		\end{pmatrix}
		=
		\begin{pmatrix}
			\dfrac{\hbar^2}{\ell^2}\eta & 0 \\[6pt]
			0 & L^2\eta
		\end{pmatrix},
		\qquad
		\begin{pmatrix}
			\mathcal{P} & \mathcal{Q} \\
			\mathcal{Q}^T & \mathcal{X}
		\end{pmatrix}^{-1}
		=
		\begin{pmatrix}
			\dfrac{\ell^2}{\hbar^2}\eta^{-1} & 0 \\[6pt]
			0 & \dfrac{1}{L^2}\eta^{-1}
		\end{pmatrix}.
	\end{equation}
	
	Substituting into the definition of $\Gamma$ yields:
	\begin{equation}
		\Gamma = \frac{\ell^2}{\hbar^2}\eta^{\mu\nu}\langle p_{\mu}\rangle \langle p_{\nu}\rangle + \frac{1}{L^2}\eta^{\mu\nu}\langle x_{\mu}\rangle \langle x_{\nu}\rangle.
	\end{equation}
	
	For the de Sitter signature $(1,4)$, raising indices gives:
	\begin{equation}
		\Gamma = \frac{\ell^2}{\hbar^2}\eta_{\mu\nu}\langle p^{\mu}\rangle \langle p^{\nu}\rangle + \frac{1}{L^2}\eta_{\mu\nu}\langle x^{\mu}\rangle \langle x^{\nu}\rangle.
	\end{equation}
	
	We emphasize that this simplified relation (33)-(34) describes a different family of quantum states (those with $\mathcal{Q}=0$) than the general states described by Eq.~(30). The general geometric equation (30) holds for all states saturating the uncertainty relations, while the simplified relation (33) applies only to the subclass with vanishing position-momentum correlations.
	
	\subsection{Identification of $L$ with the de Sitter radius and the classical limit}
	
	We may now suppose that $L = R_{ds}$ and $\ell = \ell_p$ and for $\ell \to 0$ the quantum phase space reduces to a de Sitter space with the equation
	\begin{equation}
		\frac{1}{R_{ds}^2}\eta_{\mu\nu}\langle x^{\mu}\rangle \langle x^{\nu}\rangle = -1
		\quad\Longrightarrow\quad
		\eta_{\mu\nu}\langle x^{\mu}\rangle \langle x^{\nu}\rangle = -R_{ds}^2.
	\end{equation}
	
	This is precisely the equation of de Sitter spacetime, showing that in the limit where the Planck length goes to zero, the quantum phase space reduces to a classical de Sitter geometry with curvature radius $R_{ds}$.
	
	If we normalize the invariant by setting $\Gamma = -1$, the full quantum phase space equation becomes
	\begin{equation}
		\frac{\ell_p^2}{\hbar^2}\eta_{\mu\nu}\langle p^{\mu}\rangle \langle p^{\nu}\rangle + \frac{1}{R_{ds}^2}\eta_{\mu\nu}\langle x^{\mu}\rangle \langle x^{\nu}\rangle = -1.
	\end{equation}
	
	This equation explicitly shows how the Planck length $\ell_p$ and the de Sitter radius $R_{ds}$ serve as the fundamental curvature scales for momenta space and coordinates space, respectively.
	
	\subsection{Emergence of rest mass and proper time}
	
	A profound physical interpretation arises when we consider the combination of the two fundamental limits. From the expressions in the $\mathcal{F}_0$ frame, we can construct two quantities that behave like ``squared norms'' in the fifth dimension:
	\begin{equation}
		\begin{cases}
			\langle x_4\rangle^2 - L^2 = -L^2 + L^2 = 0,\\
			\langle p_4\rangle^2 - \left(\dfrac{\hbar}{2\ell}\right)^2 = \left(\dfrac{\hbar}{2\ell}\right)^2 - \left(\dfrac{\hbar}{2\ell}\right)^2 = 0.
		\end{cases}
	\end{equation}
	
	In a generic frame, these combinations are no longer zero but instead transform under the Lorentz group $\mathrm{SO}(1,4)$ as components of a vector. This motivates the following physical identifications in the simultaneous limit where quantum gravity effects are turned off $(\ell \to 0)$ and the spacetime curvature is flattened $(L \to \infty)$:
	
	The deviation of $\langle x_4\rangle^2$ from its maximal value $L^2$ defines the square of the proper time $\tau$ of the particle:
	\begin{equation}
		\lim_{\ell \to 0, L \to \infty} \left(\langle x_4\rangle^2 - L^2\right) = c^2\tau^2.
	\end{equation}
	
	Similarly, the deviation of $\langle p_4\rangle^2$ from its minimal value $(\hbar/2\ell)^2$ defines the square of the rest mass $m$ of the particle:
	\begin{equation}
		\lim_{\ell \to 0, L \to \infty} \left(\langle p_4\rangle^2 - \left(\frac{\hbar}{2\ell}\right)^2\right) = c^2 m^2.
	\end{equation}
	
	With these identifications, the unified geometric constraint (30) reduces in the Poincaré limit to the two standard relativistic relations:
	\begin{equation}
		\langle p_0\rangle^2 - \sum_{i=1}^{3}\langle p_i\rangle^2 = c^2 m^2,\qquad
		\langle x_0\rangle^2 - \sum_{i=1}^{3}\langle x_i\rangle^2 = c^2\tau^2.
	\end{equation}
	
	The first is the familiar energy-momentum relation of special relativity, while the second expresses the invariance of the spacetime interval, with $\tau$ being the proper time experienced by the particle in its average rest frame. The symmetry group of this limit is the Poincaré group, as expected.
	
	\section{Discussion and conclusion}
	
	In this work we have examined the basic geometric structures emerging from the relativistic quantum phase space (QPS) formalism \cite{Ranaivoson2021, Ranaivoson2022, Ranaivoson2025}, with particular emphasis on the scalar invariant $\Gamma$ constructed in Sec.~2 and the geometric equation derived in Sec.~3. The QPS framework provides a geometric language in which the Planck length and the de Sitter radius appear as parameters of a single invariant relation. This suggests possible connections to quantum gravity and cosmology, but a dynamical completion is required to assess their physical realisation.
	
	The scalar invariant $\Gamma$ is built from the mean values $\langle p_{\mu}\rangle$, $\langle x_{\mu}\rangle$ and the inverse variance-covariance matrix of a quantum state. Its invariance under the symmetry group of the QPS reflects the fact that the underlying geometric structure is independent of the choice of reference frame in the combined momenta-coordinates space. For the quantum states defining the QPS that saturate by definition the uncertainty relations, i.e., the states that are as close as possible to a classical phase space point while still respecting the quantum uncertainty principle \cite{Ranaivoson2022, Ravelonjato2023, Ranaivoson2025}, the invariant takes the value $\Gamma = L^2/\ell^2$. This value is determined by the states themselves and by the identification of the scales $L$ and $\ell$; it does not depend on the reference frame. In Sec.~3 we introduced a particular reference frame $\mathcal{F}_0$ solely as a convenience: in that frame the variance-covariance matrix assumes a particularly simple block-diagonal form, and the mean values are respectively aligned along a single direction. The saturation of the uncertainty relations is an intrinsic property of the states, not of the frame, and the value of $\Gamma$ remains unchanged in any other frame related by an LCT.
	
	The central result of Sec.~3 is the basic geometric equation of the quantum phase space,
	\begin{equation}
		\begin{pmatrix}
			\langle \mathbf{p} \rangle & \langle \mathbf{x} \rangle
		\end{pmatrix}
		\begin{pmatrix}
			\mathcal{P} & \mathcal{Q} \\
			\mathcal{Q}^T & \mathcal{X}
		\end{pmatrix}^{-1}
		\begin{pmatrix}
			\langle \mathbf{p} \rangle \\
			\langle \mathbf{x} \rangle
		\end{pmatrix}
		= \frac{L^2}{\ell^2},
	\end{equation}
	which holds for states saturating the uncertainty relations. This equation combines the mean values and the quantum fluctuations in a unified manner. This is a direct manifestation of the quantum nature of the phase space: even the expectation values $\langle p_{\mu}\rangle$ and $\langle x_{\mu}\rangle$ are subject to a relation that involves the minimal and maximal uncertainties $\ell$ and $L$, and these uncertainties are always present in the state.
	
	From this general equation we can derive two physically significant limits.
	
	\emph{De Sitter spacetime limit} $(\ell \to 0)$. Taking the Planck length $\ell$ to zero corresponds to removing the minimal coordinate uncertainty, i.e., turning off the quantum-gravity scale. In this limit the maximal momentum variance diverges, while the geometric equation (41) reduces to
	\[
	\eta_{\mu\nu}\langle x^{\mu}\rangle \langle x^{\nu}\rangle = -L^2,
	\]
	because the terms containing $\langle p\rangle$ are multiplied by $\ell$ or $\ell^2$ and vanish. Thus the mean coordinates alone satisfy the de Sitter spacetime equation. The momentum degrees of freedom are not ``frozen''; rather, their constraints from the unified equation disappear, leaving only the constraint on the mean coordinates. The remaining structure is a classical curved spacetime with curvature radius $L$. The appearance of the de Sitter radius $L$ as the characteristic length scale suggests that the positive cosmological constant observed in cosmology \cite{Schneider2023, Heavens2008, LopezCorredoira2017, GenovaSantos2020, Einstein1917, deSitter1917} may have a geometric origin rooted in the quantum phase space \cite{Ranaivoson2021}.
	
	\emph{Momenta space de Sitter-like limit} $(L \to \infty)$. In the opposite regime, the de Sitter spacetime radius $L$ is sent to infinity, which corresponds to flattening the spatial curvature (zero cosmological constant). Here the maximum values of coordinate variances diverge, and the geometric equation (41) becomes:
	\[
	\eta_{\mu\nu}\langle p^{\mu}\rangle \langle p^{\nu}\rangle = -\left(\frac{\hbar}{2\ell}\right)^2,
	\]
	since the terms containing $\langle x\rangle$ are suppressed by factors $1/L^2$. This is a de Sitter-like structure in momentum space, with curvature scale $\hbar/(2\ell)$. In this limit the constraints on the mean positions are removed, leaving a curved momentum space governed by the Planck length. Such a curved momentum space is reminiscent of the structures appearing in doubly special relativity \cite{AmelinoCamelia2002, Magueijo2002} and in approaches based on the Born reciprocity principle \cite{Born1949, CastroPerelman2025}.
	
	The two limits are complementary: $\ell \to 0$ yields a classical spacetime geometry with no quantum-gravity effects, while $L \to \infty$ yields a curved momentum space with a vanishing cosmological constant. Both limits are obtained from the same unified equation (41) and highlight the reciprocity between position and momentum spaces that is inherent to the QPS formalism.
	
	This reciprocity finds a natural conceptual home in the principle of Born reciprocity, which asserts that coordinates and momenta should be treated on an equal footing in a fundamental theory \cite{Born1949, CastroPerelman2025}. In the QPS framework, the symplectic structure and the invariance under LCTs already enforce such a duality. The appearance of two fundamental scales $L$ associated with spacetime curvature and $\ell$ associated with momenta-space curvature further illustrates this reciprocity: the transformation $(L, \ell) \leftrightarrow (\hbar/(2\ell), \hbar/(2L))$ maps the de Sitter spacetime limit into the momenta-space de Sitter-like limit. This symmetric treatment of coordinates and momenta spaces is a direct consequence of the underlying quantum phase space geometry. The Born reciprocity principle thus provides a conceptual motivation for the QPS approach, and the present results demonstrate how it manifests concretely in the geometric invariants and the asymptotic regimes.
	
	A noteworthy implication of the formalism concerns the interpretation of the geometric equation (41). Unlike in classical geometry, the variables $\langle x^{\mu}\rangle$ and $\langle p^{\mu}\rangle$ are not the exact coordinates and momenta of a point particle; they are expectation values taken with respect to quantum states that necessarily possess non-zero variances. The presence of the variance-covariance matrix in the equation ensures that the quantum fluctuations are consistently accounted for. In this sense, the QPS approach provides a framework in which the classical notions of spacetime and momentum space emerge as limiting cases of a more fundamental quantum structure, where the uncertainties themselves become part of the geometric description.
	
	The results of Secs.~2 and 3 also connect naturally with the earlier work on particle physics in the same formalism \cite{Andriambololona2021, Ranaivoson2021, Ranaivoson2022, Ranaivoson2025, Randriantsoa2025}. For the signature $(1,4)$, the LCT group is isomorphic to $\mathrm{Sp}(2,8)$, whose spin representation has been shown to lead to a novel classification of the quarks and leptons which contain right-handed sterile neutrino states. Because the geometric invariant $\Gamma$ and the geometric equation are derived from the same LCT symmetry group $\mathrm{Sp}(2,8)$ that in Refs.~\cite{Andriambololona2021, Ranaivoson2025, Randriantsoa2025} was shown to predict sterile neutrinos, the present analysis is fully compatible with that particle-physics result, although it does not directly address it. This suggests that the very same geometric structures that encode the de Sitter radius and the Planck length may also govern the emergence of sterile neutrinos. It follows that these fundamental scale parameters may be relevant for understanding the origin of neutrino masses. Such connections warrant further investigation, but they already indicate that the QPS formalism offers a unified language for addressing questions in cosmology, quantum gravity, and particle physics.
	
	Other directions for future research may emerge from the results presented here. First, the geometric equation (41) was derived for states saturating the uncertainty relations corresponding to a specific choice of the invariant $\Gamma$. Extending this analysis to more general quantum states, and incorporating dynamics, could lead to a formulation of quantum field theory with a unified description of interactions on the quantum phase space, where the variance-covariance matrix acts as an additional dynamical variable. Second, the representation theory of the LCT group, and in particular the computation of its Casimir operators \cite{Randriantsoa2025}, can be further developed to classify the elementary excitations (including sterile neutrinos and description of multiple generations of quarks and leptons) that are compatible with the symmetry. Third, the two limits studied here $\ell \to 0$ and $L \to \infty$ exhibit a structural resemblance to the contraction procedures that give rise to the Carroll and Galilean groups \cite{LevyLeblond1965, LevyLeblond2022, LevyLeblond2023}. Investigating whether such a correspondence can be made precise within the QPS framework may provide a deeper understanding of how classical and quantum regimes are connected through the geometry of phase space. The idea of incorporating quantum fluctuations into geometry (Feynman) and the quantum origin of spacetime (Hawking) resonates with our approach, although the present work does not implement a full quantum gravity theory \cite{Feynman1948, Hawking1984}.

	\section*{Acknowledgments}
	
	The authors would like to express their sincere gratitude to the Institut National des Sciences et Techniques Nucléaires (INSTN-Madagascar) for providing a stimulating research environment. Special thanks are extended to our colleagues for fruitful discussions and valuable feedback during the preparation of this work.
	
	\section*{Declarations}
	\noindent\textbf{Data Availability:} No new data were created or analyzed in this study. Data sharing is therefore not applicable to this article.\\[0.2cm]
	\noindent\textbf{Ethics Approval:} Not applicable. This study is purely theoretical and does not involve human participants, animals, or human-derived materials.\\[0.2cm]
	\noindent\textbf{Funding:} This research received no external funding.\\[0.2cm]
	\noindent\textbf{Conflict of interest:} The authors declare no conflict of interest.
	\noindent\textbf{Declaration of generative AI and AI-assisted technologies in the writing process:}
	The author(s) utilized generative AI solely for translating specific sections and improving the English phrasing to a native level. All outputs were critically examined and approved by the author(s), who assume full responsibility for the work.
	
	\bibliographystyle{apsrev4-1}

\begin{thebibliography}{99}
		
		\bibitem{Ranaivoson2021}
		R.T. Ranaivoson, Raoelina Andriambololona, H. Rakotoson, R. Raboanary, Linear Canonical Transformations in Relativistic Quantum Physics, Phys. Scr. 96, 065204 (2021), DOI: 10.1088/1402-4896/abeba5.
		.
		
		\bibitem{Ranaivoson2022}
		R.T. Ranaivoson, Raoelina Andriambololona, H. Rakotoson, R.H.M. Ravelonjato, Invariant quadratic operators associated with Linear Canonical Transformations and their eigenstates, J. Phys. Commun. 6, 095010 (2022), DOI: 10.1088/2399-6528/ac8520.
		.
		
		\bibitem{Ravelonjato2023}
		R.H.M. Ravelonjato, R.T. Ranaivoson, Raoelina Andriambololona, R. Raboanary, H. Rakotoson, N. Rabesiranana, Quantum and Relativistic Corrections to Maxwell-Boltzmann Ideal Gas Model from a Quantum Phase Space Approach, Found. Phys. 53, 88 (2023), DOI: 10.1007/s10701-023-00727-5.
		
		\bibitem{Andriambololona2017}
		Raoelina Andriambololona, R.T. Ranaivoson, H.D. Randriamisy, H. Rakotoson, Dispersion Operators Algebra and Linear Canonical Transformations, Int. J. Theor. Phys. 56, 1258-1273 (2017), DOI: 10.1007/s10773-016-3268-4.
		
		\bibitem{Andriambololona2021}
		Raoelina Andriambololona, R.T. Ranaivoson, H. Rakotoson, R. Raboanary, Sterile neutrino existence suggested from LCT covariance, J. Phys. Commun. 5, 091001 (2021), DOI: 10.1088/2399-6528/ac2409.
		
		\bibitem{Ranaivoson2025}
		R.T. Ranaivoson, Raoelina Andriambololona, H. Rakotoson, R. Raboanary, J. Rajaobelison, P.M. Randriantsoa, Quantum Phase Space Symmetry and Sterile Neutrinos, J. Subat. Part. Cosmol. 3, 100039 (2025), DOI: 10.1016/j.jspc.2025.100039.
		
		\bibitem{Randriantsoa2025}
		P.M. Randriantsoa, R.T. Ranaivoson, Raoelina Andriambololona, R. Raboanary, W.C. Solofoarisina, A.F.H. Rasaminanana, Casimir operators for the relativistic quantum phase space symmetry group, arXiv:2512.18262 [quant-ph], (2025), DOI: 10.48550/arXiv.2512.18262.
		.
		
		\bibitem{Schneider2023}
		Mike D. Schneider, Empty space and the (positive) cosmological constant, Studies in History and Philosophy of Science, 100, 12-21, (2023), DOI: 10.1016/j.shpsa.2023.05.008.
		
		\bibitem{Heavens2008}
		A. Heavens, The cosmological model: an overview and an outlook, J. Phys.: Conf. Ser. (2008), DOI: DOI 10.1088/1742-6596/120/2/022001.
		
		\bibitem{LopezCorredoira2017}
		Martin Lopez-Corredoira, Tests and problems of the standard model in Cosmology, Found Phys 47, 711-768 (2017), DOI: 10.1007/s10701-017-0073-8.
		
		\bibitem{GenovaSantos2020}
		Ricardo T. Genova-Santos, The establishment of the Standard Cosmological Model through observations, In: Kabath, P., Jones, D., Skarka, M. (eds) Reviews in Frontiers of Modern Astrophysics. Springer, Cham (2020), DOI: $10.1007/978-3-030-38509-5_11$.
		
				
		\bibitem{Andriambololona1985}
		Raoelina Andriambololona, Alg\`{e}bre lin\'{e}aire et multilin\'{e}aire, Collection LIRA (1985).
		
		
		\bibitem{Meschini2007}
		D. Meschini, Planck-scale physics: facts and beliefs, Found Sci 12, 277–294 (2007), DOI: 10.1007/s10699-006-9102-3.
		
		\bibitem{Einstein1917}
		A. Einstein, Kosmologische Betrachtungen zur allgemeinen Relativitätstheorie, Sitzungsber. Preuss. Akad. Wiss. 1, 142-152 (1917).
		
		\bibitem{deSitter1917}
		Friedman, A. On the Curvature of Space. General Relativity and Gravitation 31, 1991–2000 (1999). https://doi.org/10.1023/A:1026751225741.
		
		\bibitem{AmelinoCamelia2002}
		G. Amelino-Camelia, Relativity in spacetimes with short-distance structure governed by an observer-independent (Planckian) length scale, Int. J. Mod. Phys. D 11, 35 (2002), DOI: 10.1142/S0218271802001330.
		
		\bibitem{Magueijo2002}
		J. Magueijo, L. Smolin, Lorentz invariance with an invariant energy scale, Phys. Rev. Lett. 88, 190403 (2002), DOI: 10.1103/PhysRevLett.88.190403.
		
		\bibitem{Born1949}
		M. Born, Reciprocity theory of elementary particles, Rev. Mod. Phys. 21, 463 – Published 1 July, 1949, DOI: 10.1103/RevModPhys.21.463.
		
		\bibitem{CastroPerelman2025}
		C. Castro Perelman, Born Reciprocal (Non-inertial) Relativity, Phase Space Trajectories and Strings with variable Tension, Mod. Phys. Lett. A, DOI: 10.1142/S0217732326500781.
		
		\bibitem{LevyLeblond1965}
		J.-M. L\'{e}vy-Leblond, Une nouvelle limite non-relativiste du groupe de Poincar\'{e}, Annales de l'institut Henri Poincaré. Section A, Physique Théorique, Tome 3 (1965) no. 1, pp. 1-12.
		
		\bibitem{LevyLeblond2022}
		J.-M. L\'{e}vy-Leblond, On the unexpected fate of scientific ideas. An archeology of the Carroll group, in: 34th International Colloquium on Group Theoretical Methods in Physics, Strasbourg, 18-22 July 2022.
		
		\bibitem{LevyLeblond2023}
		J.-M. L\'{e}vy-Leblond, Quand Galil\'{e}e et Carroll engendrent Lorentz, Ann. Henri Poincar\'{e} 24, 3209 (2023).
		
		\bibitem{Feynman1948}
		R.P. Feynman, Space-Time Approach to Non-Relativistic Quantum Mechanics, Rev. Mod. Phys. 20, 367 (1948), DOI: 10.1103/RevModPhys.20.367.
		
		\bibitem{Hawking1984}
		S.W. Hawking, The quantum state of the universe, Nucl. Phys. B 239, 257 (1984), DOI: 10.1016/0550-3213(84)90093-2.
		
	\end{thebibliography}

\end{document}